\documentclass{emulateapj}

\usepackage{natbib}
\newcommand{\CLA}{{\mbox{\tiny CLA}}}
\newcommand{\eccentricity}{e}

\newcommand{\GW}{{\mbox{\tiny GW}}}

\newcommand{\KMS}{\mbox{km s}^{-1}\,}

\newcommand{\MIN}{{\mbox{\tiny min}}}
\newcommand{\QUASICIRCULAR}{{\mbox{\tiny qc}}}

\newcommand{\TOTAL}{{\mbox{\tiny total}}}

\begin{document}
%

\title{Gravitational recoil velocities from eccentric binary black hole mergers}

\author{Carlos F. Sopuerta\altaffilmark{1}, Nicol\'as Yunes\altaffilmark{2}, and Pablo Laguna\altaffilmark{2}}

\affil{Center for Gravitational Wave Physics, \\
The Pennsylvania State University, 
University Park, PA 16802, USA}

\altaffiltext{1}{Department of Physics, University of Guelph, Guelph, Ontario, Canada N1G 2W1}

\altaffiltext{2}{IGPG, Departments of Astronomy \& Astrophysics and Physics, 
The Pennsylvania State University, University Park, PA 16802, USA}


\begin{abstract}
  The formation and growth of supermassive black holes is a key issue
  to unveil the secrets of galaxy formation.  In particular, the
  gravitational recoil produced in the merger of unequal mass black
  hole binaries could have a number of astrophysical implications,
  such as the ejection of black holes from the host galaxy or globular
  cluster.  We present estimates of the recoil velocity that include
  the effect of small eccentricities. The approach is specially suited
  for the last stage of the merger, where most of the emission of
  linear momentum in gravitational waves takes place.  Supplementing
  our estimates with post-Newtonian approximations, we obtain lower
  and upper bounds that constrain previous recoil velocities estimates
  as well as a best estimate that agrees with numerical simulations in
  the quasi-circular case.  For eccentricities $e \le 0.1$, the
  maximum recoil is found for mass ratios of $M_1/M_2 \sim 0.38$ with
  a best estimate of $\sim 167\, (1 + \eccentricity),\KMS$ and upper and lower bounds of
  $79\, (1 + \eccentricity)\,\KMS$ and $216 \, (1 + \eccentricity) \,\KMS$ respectively.
\end{abstract}

\keywords{
          black hole physics ---
          cosmology: theory ---
          galaxies: nuclei ---
          gravitation ---
          gravitational waves ---
          relativity
}

\section{Introduction}\label{intro}
In present hierarchical cold dark matter cosmologies, large-scale
structures are formed by the merger of small (sub-galactic) structures
that originated from small primordial density
perturbations~\citep{Press:1974ps}.  In the case of galaxies, there is
evidence that most of the nearby ones host super-massive black holes
(SMBH) at their centers~\citep{Richstone:1998ea,Magorrian:1997hw} with
masses in the range $10^5 - 10^9 M^{}_{\odot}$.  Moreover,
observations have revealed tight relations between the SMBH and the
bulge of the host galaxy~\citep{Ferrarese:2000fm}, and indicate also
that the SMBH mass may be determined by the mass of the host dark
matter halo~\citep{Ferrarese:2002la}.  This suggests that there must
be a deep relation between the formation mechanism of the SMBH and the
host galaxy which is not yet completely understood.  It has been
suggested~\citep{Volonteri:2002am} that SMBHs grow as a combination of
gas accretion and mergers with other SMBHs that come together as a
result of the merger of their host dark matter halos and dynamical
friction~\citep{Begelman:1980vb}.

The last stages of the merger of SMBHs will be driven by the emission
of gravitational radiation in the low-frequency band.  These
extremely energetic events will be observable by the planned
space-based gravitational wave antenna
LISA~\citep{Danzmann:2003ad,Prince:2003aa}.  In addition to energy and
angular momentum, in the case of unequal mass SMBHs, there is a net
flux of linear momentum carried away form the system by the
gravitational waves~\citep{Peres:1962ap,Bekenstein:1973jd}.  Then,
momentum conservation implies that the final SMBH remnant will
experience a recoil.  There may be observational evidence of an
ejected SMBH from an ongoing galaxy merger~\citep{Haehnelt:2006hd}
either from gravitational recoil or gravitational slingshot of three o
more SMBHs in the merger.  The knowledge of the magnitude of the
recoil velocity is crucial to understand the demography of SMBHs at
the centers of galaxies and in the interstellar and intergalactic
media, and their apparent absence in dwarf galaxies and stellar
clusters~\citep{Madau:2004mq,Merritt:2004xa}.  An estimate of the
recoil can also be used to constrain theories in which SMBHs grow at
the center of dark matter halos~\citep{Haiman:2004ve} and to estimate
SMBH merger rates~\citep{Micic:2006ta}.

In this letter, we present the results of a calculation of the recoil
velocity based on an approximation scheme which is valid when the
black holes (BHs) are very close to each other, that is, in the last 
stage of the binary merger. It is during this plunge  
where most of the recoil is accumulated.  This contrasts with 
previous calculations that use approximations valid when the BHs are well-separated,
except for recent fully numerical relativistic 
calculations~\citep{Baker:2006vn,Herrmann:2006ks,Gonzalez:2006md}).  We include the
effect of eccentricity in the recoil velocity since there are some
indications that SMBH binaries may have not completely circularized by
the time of merger~\citep{Aarseth:2003aa,Armitage:2005an,Dotti:2006dc}.

\section{Recoil velocities and the Close Limit approximation} \label{resultscla}
General relativity predicts that an unequal mass binary system in
coalescence produces anisotropic emission of
gravitational waves that carry linear momentum away from the binary. 
As a consequence, the center of mass of the system experiences a recoil 
velocity~$v^{}_{r}$~\citep{Wiseman:1992dv}.  The first estimates, done
by~\citet{Fitchett:1983fc} using a quasi-Newtonian approach, yielded a
maximum $v^{}_r \sim 1480\,\KMS (r^{}_{g}/r^{}_{f})^4$ for a symmetric
mass ratio of $\eta = M^{}_{1} M^{}_{2}/M^2 \sim 0.2\,$, where
$r^{}_{g} = 2\,GM/c^2$ is the gravitational radius, $M^{}_{1,2}$ and
$M$ are the individual and total masses respectively, and
$r^{}_{f}$ is the orbital separation at which gravitational-wave
emission ends.  Similar results up to the innermost stable circular
orbit (ISCO) where found using a relativistic perturbative
method~\citep{Fitchett:1984fd}.  This velocity estimate is large
compared to galactic escape velocities.  In contrast, the
first post-Newtonian (PN) calculation~\citep{Wiseman:1992dv} for
neutron star binaries produced a much lower result: $v^{}_r <
1\,\KMS$.

Motivated by the astrophysical impact of the gravitational recoil,
there has been recently a number of efforts to obtain better estimates of
the recoil velocity.  Using a relativistic perturbative scheme for the
description of extreme-mass ratio ($\eta \ll 1$) binaries and
including gravitational back-reaction,~\citet{Favata:2004wz} and \citet{Merritt:2004xa} 
obtained maximum recoil at $\eta \sim 0.2\,$ with 
$167\,\KMS \le v^{}_{r} \le 622 \,\KMS$, depending on the spin parameter of
the massive black hole ($v^{}_r = 456\,\KMS$ for the non-spinning case). Recently,
\citet{Blanchet:2005rj} used second-order PN
approximations up to ISCO, followed by an integration of the 2PN
linear-momentum flux along a {\it plunge} geodesic,
to obtain a maximum kick velocity $v^{}_{r} = 250\pm 50\,\KMS$
for a symmetric mass ratio $\eta = 0.2\,$.  On the other hand, estimates with the
effective one body approach~\citep{Damour:2006tr} 
give a maximum recoil in the range $49-172\,\KMS$ for the same $\eta = 0.2$.
Using the hybrid perturbative-fully relativistic Lazarus
approach,~\citet{Campanelli:2004zw} obtained also that the maximum recoil takes place
at $\eta \sim 0.2$ with a magnitude of 
$\sim 250\,\KMS$.
Currently, there are very few fully general relativistic numerical simulations of kicks.
\citet{Baker:2006vn} reported $v^{}_r = 105\,\KMS$ for a symmetric mass ratio 
$\eta = 0.24$, while \citet{Herrmann:2006ks} obtained $v^{}_r = 82\,\KMS$ for
$\eta = 0.23$. Recently,~\citet{Gonzalez:2006md} report a maximum recoil
$v^{}_r = 176 \pm 11\,\KMS$ for a symmetric mass ratio of $\eta = 0.195\pm 0.005$,
where the error bars account only for finite differencing Einstein's equations.

Except for the fully relativistic approaches, which have their own 
uncertainties (mainly from computational resource limitations), all 
other approaches become less accurate the closer the BHs
get.  Motivated by the fact that the main contribution to the recoil
comes from the last stage of the merger and plunge, as many of the
previously cited studies suggest, in this letter we compute the
distribution of kicks using a scheme that works better the closer the
orbital separation.  This scheme~\citep{Price:1994pm}, known as the
close-limit approximation (CLA), has been shown to provide accurate
results for head-on collisions when compared to full numerical
relativity~\citep{Anninos:1995vf}.  The CLA is based on the idea that
the last stages of binary black hole (BBH) mergers can be modeled as a
single perturbed BH.
Gravitational recoil was studied in the context of the CLA scheme for the case of
head-on collisions of BHs starting from rest
by~\citet{Andrade:1996pc}.  Recently, we \citep{Sopuerta:2006wj}
extended this analysis to the quasi-circular case and 
found a maximum recoil of
$v^{}_{r}\sim 57\,\KMS$ for $\eta \sim 0.19$ starting from a
separation of $4\,GM/c^2$, the maximum separation at which it has been
estimated that the CLA scheme works~\citep{Gleiser:1996yc}.  This
estimate covers the last part of the merger, plunge and ringdown.  One
can complement this result with PN estimates from the inspiral phase.
If we consider the contribution from PN approximations up to ISCO, we
obtain a lower bound for the maximum recoil of $v^{}_{r}\sim
80\,\KMS\,$, whereas if we push the PN approximations up to the point where we start the
CLA scheme, we get an upper limit in the maximum recoil of $v^{}_{r}\sim 215\,\KMS$
(it is known that PN and perturbative schemes overestimate the
linear-momentum flux in the strong field region).  Alternatively, if,
instead of pushing the PN method towards small separations, we push the
CLA scheme towards larger separations around $5\,GM/c^2$, we get a
slightly larger upper bound.

In this letter, we improve and extend these estimates for initial
configurations corresponding to BBHs in eccentric orbits.  The most
important point in the implementation of the CLA scheme is to
establish a correspondence between the initial configuration
representing a BBH and a perturbed single BH.  This correspondence
allows us to identify, in the initial BBH configuration, a background
(a non-spinning Schwarzschild BH) and perturbative multipolar
gravitational modes~\citep{Sopuerta:2006wj}.  We then evolve these
modes by using the machinery of BH perturbation theory and, from the
results of the evolution, we can compute the fluxes of energy, angular
momentum, and linear momentum ($\dot{P}^i_{\GW}$) emitted by the BBH
system during the merger.  Then, the recoil velocity is given by
\begin{equation}
v^{}_r = \|v^i_r\|\,,\qquad
v^i_r= - M^{-1} \int^{t^{}_f}_{t^{}_i} dt\,\dot{P}^i_{\GW}\,.
\end{equation}
where $t^{}_i$ is the time corresponding to
the initial separation of the BBH, and $t^{}_f$ is some time after
merger, when the gravitational wave emission becomes negligible.  
In our calculations, we use an initial BBH configuration~\citep{Brill:1963bl,Bowen:1980yu}
that has been shown to accurately represent binaries to Newtonian order.  They are
characterized by four parameters~\citep{Sopuerta:2006wj}: the total
mass of the system ($M$), the symmetric mass ratio of the binary
($\eta$), the initial separation ($d$), and the initial
linear momentum ($P$) of each BH.  
$M$ and $\eta$ enter as a scale and input parameters, respectively.
Given $\eta$, the type of orbit is determined by the pair
$(d,P)$.  For a quasi-circular orbit we find a relation between $P$ and
$d$ by minimizing the gravitational binding energy of the black hole
binary.  This relation turns out to be analogous to the Newtonian one,
although it has to be interpreted within the framework of general
relativity.  

In this work, in order to study the effect of eccentricity in the
gravitational recoil, we take the distance parameter to be $d = a (1 -
\eccentricity)$, where $a$ is a parameter distance that plays the role
of the semi-major axis and $\eccentricity$ is the eccentricity.  Then,
the linear momentum parameter is given by
\begin{equation}
P =  P^{}_{\QUASICIRCULAR}(a)\,\sqrt{\frac{1+\eccentricity}{1-\eccentricity}} \,, \label{P}
\end{equation}
where $P^{}_{\QUASICIRCULAR}(a)$ is the linear momentum parameter for
quasi-circular configurations~\citep{Sopuerta:2006wj} with radius $a$.

\begin{figure}
\plotone{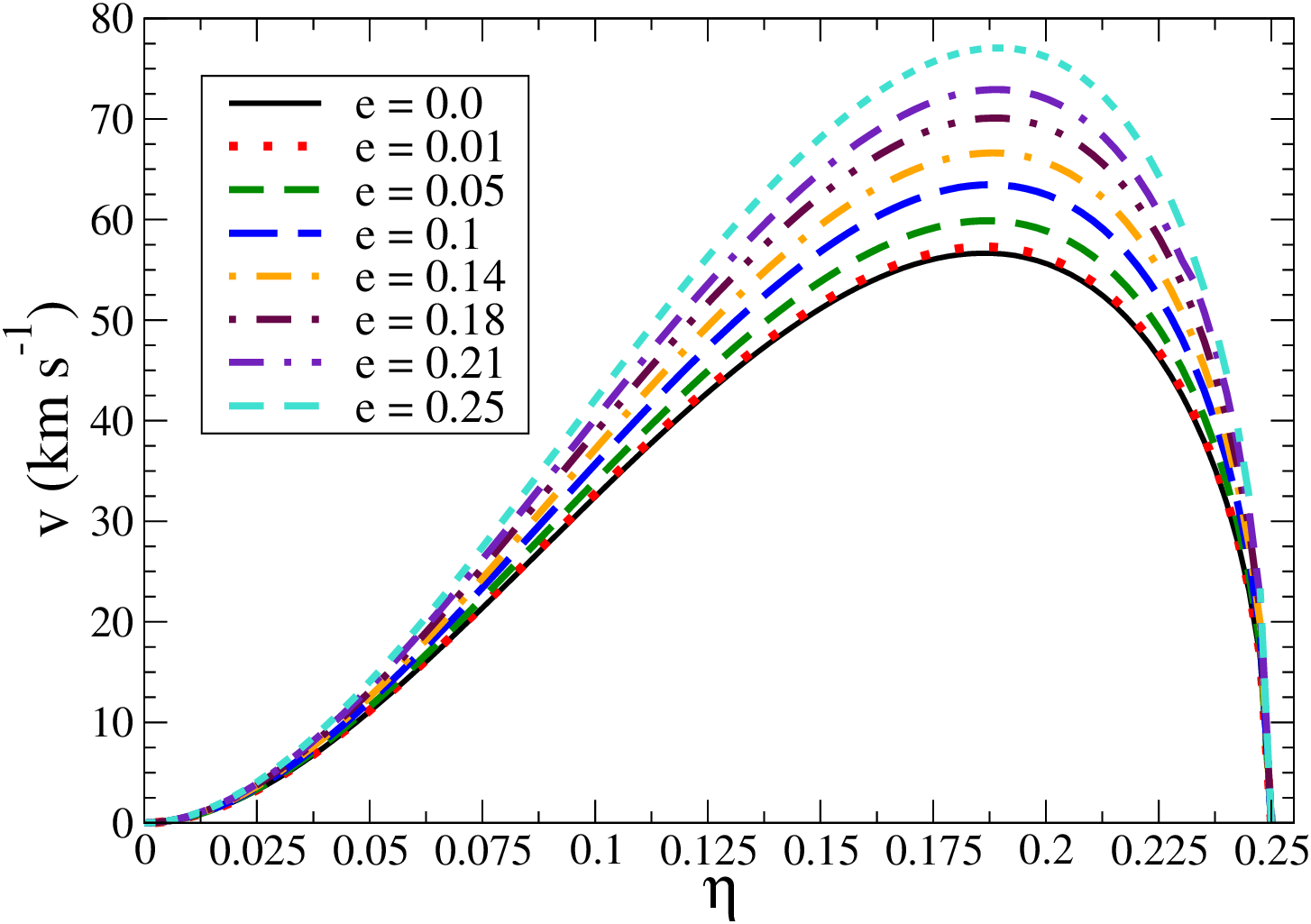}
\caption{\label{cla_v}Recoil velocity in $\KMS$ as a function of
  symmetric mass ratio $\eta$ for eccentricities in the range $\eccentricity = 0-0.25\,$.}
\end{figure}

For our kick estimates, we have kept the semi-major axis
fixed ($a = 4\,GM/c^2$) and only varied the symmetric mass ratio and
eccentricity parameters $(\eta,\eccentricity)$.  Figure~\ref{cla_v}
shows the recoil velocity $v^{}_r$
as a function of the symmetric mass ratio $\eta$ for several values of
the eccentricity: $\eccentricity = 0 - 0.25\,$.
We observe that the maximum recoil occurs at  $\eta \sim 0.19$,
consistent with previous results.
As we increase the
eccentricity, the recoil velocity also increases by approximately a factor of
$1 + \eccentricity$ for eccentricities in the range $\eccentricity < 0.1$.
This increase can be qualitatively understood by expanding equation~(\ref{P})
for small eccentricities, which leads to an increase by a factor $1 + \eccentricity$
in the linear momentum parameter.  Thus, it seems that an increase in the
plunge velocity leads to roughly the same increase in the recoil velocity.
If we consider the entire range of eccentricities that we considered, 
the numerical results can be fitted to the following
two-parameter non-linear function
\begin{eqnarray}
v^{\CLA}_r(\eta,\eccentricity) &=& \alpha \; \sqrt{1 - 4 \eta} \;\eta^2 \left(1
+ \beta^{}_1\; \eta + \beta^{}_2 \; \eta^2 \right)
\nonumber \\
& \times &
\frac{1 + \eccentricity}{1 - \eccentricity} \left(1 + \gamma^{}_1\; \eccentricity + \gamma^{}_2 \;
\eccentricity^2   + \gamma^{}_3 \; \eccentricity^3\right)\,,\label{fit_eq}
\end{eqnarray}
where $\alpha$ ($\KMS$), $\beta^{}_1$, $\beta^{}_2$, $\gamma^{}_1$, $\gamma^{}_2$, and
$\gamma^{}_3$ are fitting parameters given in Table~\ref{table_fit} in the CLA row.
The fit was performed using numerical results in the range $\eta = 0-0.25$ and
$\eccentricity = 0 - 0.5$ with an average error $< 1\,\KMS\,$.
The fitting function~(\ref{fit_eq}) can be shown to be functionally equivalent to that
presented by~\citet{Fitchett:1983fc} for small eccentricities, but with
different coefficients that lead to significantly smaller magnitudes for the recoil.

\section{Estimating the total recoil} \label{totalrecoil}
The estimates of the recoil velocities for BBH mergers provided by
equation~(\ref{fit_eq}) and the first row of Table~\ref{table_fit} are
incomplete because we have not taken into account the contribution
from the inspiral ($\sim 20\,\KMS$) and the beginning of the merger,
where fully non-linear computations are needed.  In order to obtain an
estimate of the total accumulated recoil velocity, we can supplement
our calculations with estimates from approximation techniques that are
accurate for moderate and large separations. In this letter, we use a
PN approximation that provides the accumulated kick up to some minimum
separation. For quasi-circular orbits~\citep{Blanchet:2005rj} and
assuming the angular frequency is given by the Newtonian estimate
$\omega = \sqrt{M/d^3}$, the accumulated recoil up to some minimum
separations of $4\,GM/c^2$ and $6\,GM/c^2$ respectively can be fitted
by Eq.~(\ref{fit_eq}) with the second and third rows of
Table~\ref{table_fit} respectively. As in~\citet{Sopuerta:2006wj},
upper and lower limits can be obtained by combining these PN estimates
with the CLA calculation. 

\begin{figure}
\plotone{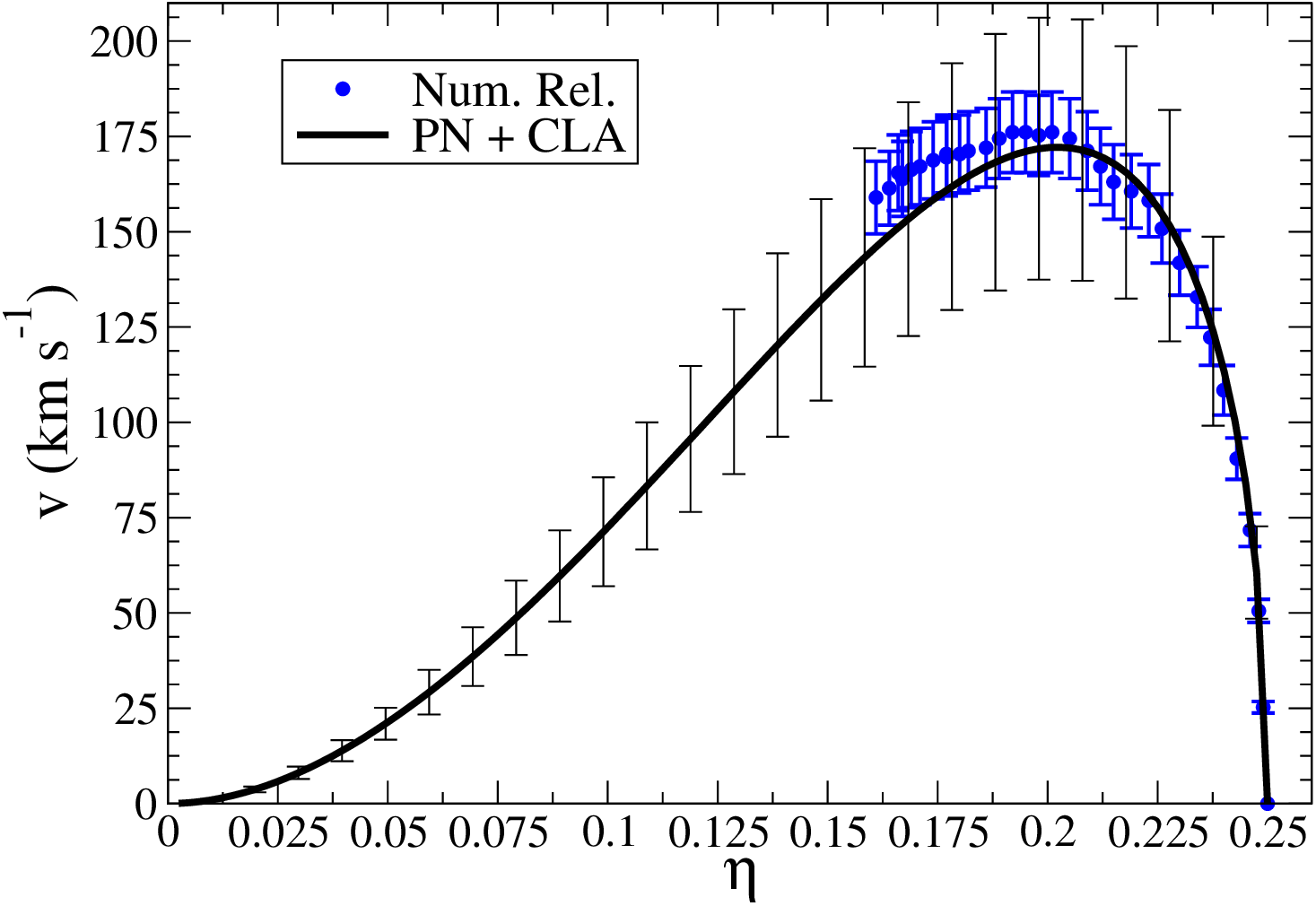}
\caption{\label{jena} Best estimate of the recoil velocity in $\KMS$ as a function of
  symmetric mass ratio $\eta$ for the quasi-circular case
  $\eccentricity = 0\,$, calculated by combining the CLA calculation
  with a $2$ PN approximation. Also plotted are the results from a
  full numerical relativistic simulation~\citep{Gonzalez:2006md}.}
\end{figure}

Our best estimate consists in adding the PN accumulated recoil from
infinite to a separation of $4\,GM/c^2$ to the CLA estimates, obtained
by starting the CLA at that separation.  However, to compensate for
using the PN approximation in a regime where it becomes less accurate,
we use a PN expression for the angular velocity with errors of
${\cal{O}}(v)^5$, instead of the Newtonian value used
in~\citet{Sopuerta:2006wj} to obtain an upper limit.  The results for
the quasi-circular case are shown in Fig.~\ref{jena}, with error bars
coming from the PN approximate error, and compared to the most recent
full relativistic results of~\cite{Gonzalez:2006md} with excellent
agreement.

The main caveat of supplementing the CLA with a PN approximation is
that presently the latter is only available for quasi-circular
orbits~\citep{Blanchet:2005rj}. However, by comparing the contribution from
the CLA scheme with that provided by the PN approximation for the
quasi-circular case, we find that the total recoil velocity can be
accurately described by the CLA contribution times a factor:
\begin{equation}
\label{tot_v}
v^{\TOTAL}_{r} = v^{\CLA}_{r} \left(1 + {\cal{E}} \right),
\end{equation}
so that ${\cal{E}}$ can be written in terms of the symmetric mass ratio $\eta$ as
\begin{equation}
\label{F}
{\cal{E}} = \kappa\, (1 + \lambda^{}_1 \;\eta + \lambda^{}_2 \; \eta^2),
\end{equation}
where the coefficients $\kappa$, $\lambda^{}_1$, and $\lambda^{}_2$
depend on the minimum distance that we use in the PN approximation.
When the minimum distance is taken to be $r_1^{\MIN} = 4\,GM/c^2$ we
obtain: $\kappa = 1.357\;(0.648)$, $\lambda^{}_1 = 4.418\;(5.612)$,
and $\lambda^{}_2 = 5.33\;(23.63)$ [the estimates in parenthesis
correspond to using the 2PN angular velocity instead of the Keplerian
one]; whereas when we take it to be $r_2^{\MIN} = 6\,GM/c^2$ the
coefficients are: $\kappa = 0.199$, $\lambda^{}_1 = 4.223$, and
$\lambda^{}_2 = 3.926$.

\begin{figure}
\plotone{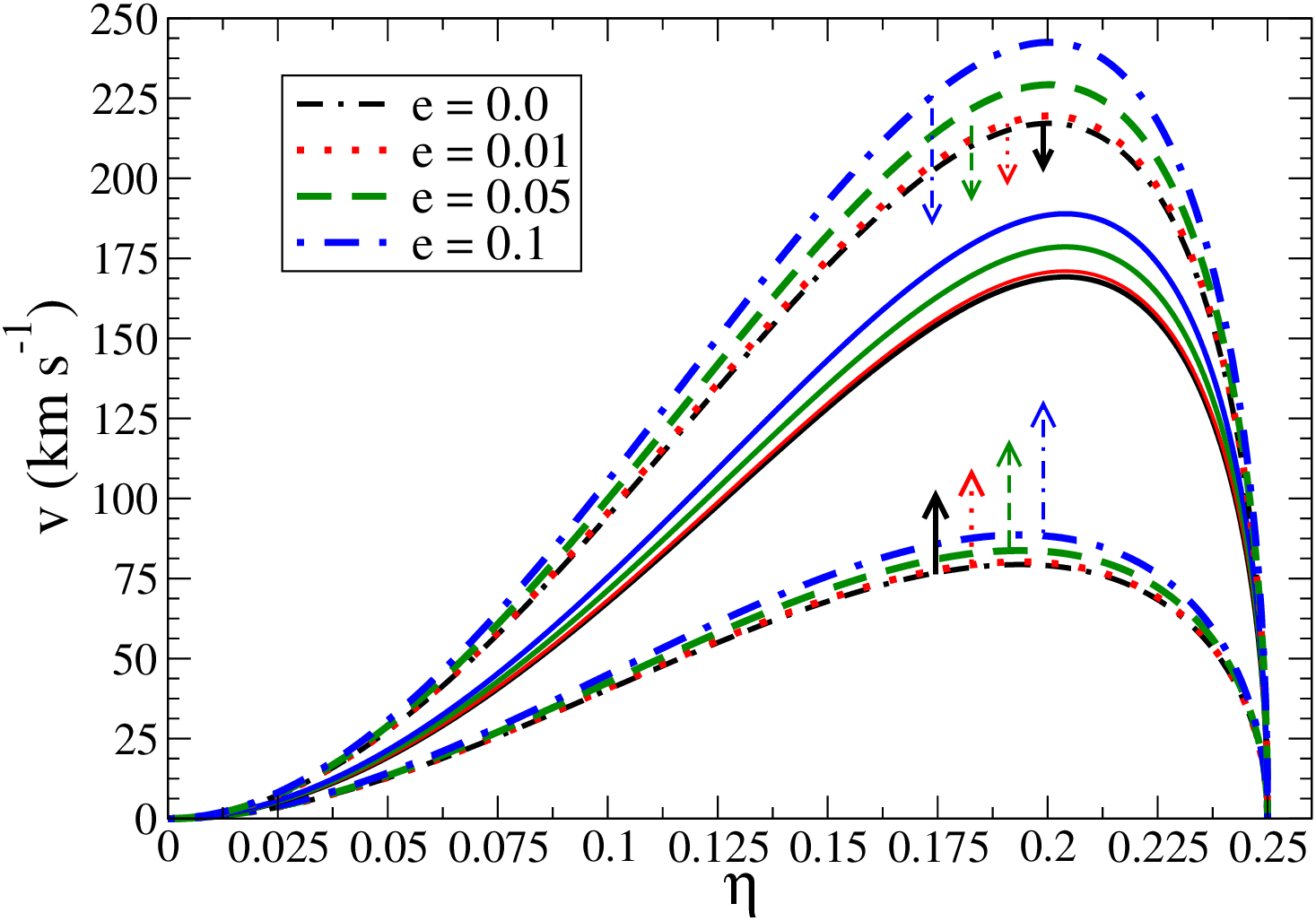}
\caption{\label{upl}Upper and lower limits (dashed and dotted lines)
  and best estimate (solid lines) to the total recoil velocity in
  $\KMS$ as a function of symmetric mass ration $\eta$ for
  eccentricities in the range $e=0-0.1\,$.}
\end{figure}

We can then obtain lower and upper bounds as well as a best estimate
for the total recoil velocities from eccentric BBH mergers if we
assume that expressions~(\ref{tot_v}) and~(\ref{F}) can be
extrapolated to the range of eccentricities under consideration,
$\eccentricity \le 0.25$.  When we use $r_2^{\MIN} = 6\,GM/c^2$ and
the Keplerian estimate for the angular velocity, the total recoil
obtained is a lower limit because we do not account for the
contribution in the region $4 - 6\,GM/c^2$.  On the other hand, if we
use $r^{\MIN}_1=4\,GM/c^2$ and the Keplerian angular velocity,
Eq.~(\ref{tot_v}) provides an upper limit since it is well-known that
the PN approximation overestimate the recoil in the $4 - 6\, GM/c^2$
range.  Our best estimate consists in complementing the CLA with a PN
estimate using $r^{\MIN}_1=4\,GM/c^2$ and the 2PN angular velocity.
These estimates, lower and upper bounds and best estimate, are shown
in Figure~\ref{upl}.

\section{Summary and Discussion}\label{discussion}
We applied the CLA scheme to the computation of recoil velocities for
BBH mergers including the effect of eccentricity.  For small
eccentricities, $\eccentricity < 0.1$, we find a generic increase in
the recoil of the order of $10\,\%$ with respect to the quasi-circular
case. This increase is related to the fact that, for slightly
eccentric orbits, the magnitude of the initial velocity of each BH
increases roughly by a factor of $1 + \eccentricity$.  Since the CLA
scheme is valid for separations smaller than the corresponding one for
the ISCO, we have supplemented our results with estimates obtained
with a $2$PN approximation, which work very well for large
separations.  Combining appropriately both approximation schemes, we
have produced lower and upper bounds, as well as a best estimate, for
the recoil velocity.

Our calculations indicate that the maximum recoil velocity takes place
at a symmetric mass ratio of $\eta \sim 0.19$ and its magnitude can be
as low as $\sim 79-88\,\KMS$ and as high as $\sim 216-242\,\KMS$, with
a best estimate in the range $\sim 167 - 187\,\KMS$.  The variation in
the upper and lower bounds is due to the eccentricity parameter of the
orbit. All our estimates can be fitted by non-linear function of
$\eta$ and $\eccentricity$ given by Eq.~(\ref{fit_eq}).  This formula
is expected to work well for small eccentricities, and it reduces to
the expression given by~\citet{Fitchett:1983fc}, with a significantly
smaller magnitude, when $\eccentricity \ll 1$. These results also
significantly narrow previous
calculations~\citep{Favata:2004wz,Blanchet:2005rj,Damour:2006tr} and
agree remarkably well with the most recent full relativistic
computations~\citep{Gonzalez:2006md} in the quasi-circular case.

One of the main conclusions that can be extracted from our and recent
estimates of the gravitational recoil from BBH mergers is that
the main contribution to the recoil comes from a narrow interval of
separations that include the ISCO.  Not surprisingly, this coincides
with the non-linear gravitational regime, where perturbative
approximation schemes tend to break down. In this sense, the
attractive feature of our approach is that it uses an approximation
method, the CLA scheme, valid for small separations, and complements
the estimate with another approximation, the PN scheme, valid for
large to moderate separations.

As mentioned before, SMBH merger recoil velocities are relevant for hierarchical
dark matter scenarios for structure formation, as well as in the understanding of
the displacement of nuclear structure in dense galaxies and the distortion
of X-shaped radio sources. Our results do not significantly alter the astrophysical
conclusions obtained previously by~\citet{Merritt:2004xa}. Instead, our
results confirm and sharpen these conclusions, by providing a narrower
lower and upper bounds for the possible recoil velocities.

Future work can assess the robustness of our
estimates by studying their dependence on 
the initial configuration as well as the inclusion of higher
gravitational multipoles in the calculation.  Moreover, given 
the recent advances in numerical
relativity, it should be possible to compare results and to determine
with better precision the distribution of recoil velocities in the
parameter space $(\eta,\eccentricity)$.  A very important issue is
the effect of spins, which has not yet been dealt with enough
generality to ascertain what is the impact of the additional scales
introduced by spinning BHs. 

\acknowledgments The authors acknowledge the support of the Center for
Gravitational Wave Physics funded by the National Science Foundation
under Cooperative Agreement PHY-0114375, and support from NSF grants
PHY0555628, PHY0555436, PHY0218750, PHY0244788, and PHY0245649.  CFS
was partially supported by the Natural Sciences and Engineering
Research Council of Canada.  The Information Technology Services at
Penn State University is also acknowledged for the use of their
computer clusters.


\begin{thebibliography}{35}
\expandafter\ifx\csname natexlab\endcsname\relax\def\natexlab#1{#1}\fi
\expandafter\ifx\csname href\endcsname\relax
  \def\href#1#2{}\fi
\expandafter\ifx\csname urllinklabel\endcsname\relax
  \def\urllinklabel{[LINK]}\fi
\expandafter\ifx\csname adsurllinklabel\endcsname\relax
  \def\adsurllinklabel{[ADS]}\fi

\bibitem[{{Aarseth}(2003)}]{Aarseth:2003aa}
{Aarseth}, S.~J. 2003, \apss, 285, 367


\bibitem[{Andrade \& Price(1997)}]{Andrade:1996pc}
Andrade, Z. \& Price, R.~H. 1997, Phys. Rev. D, 56, 6336


\bibitem[{Anninos {et~al.}(1995)Anninos, Price, Pullin, Seidel, \&
  Suen}]{Anninos:1995vf}
Anninos, P., Price, R.~H., Pullin, J., Seidel, E., \& Suen, W.-M. 1995, Phys.
  Rev. D, 52, 4462


\bibitem[{{Armitage} \& {Natarajan}(2005)}]{Armitage:2005an}
{Armitage}, P.~J. \& {Natarajan}, P. 2005, \apj, 634, 921


\bibitem[{Baker {et~al.}(2006)}]{Baker:2006vn}
Baker, J.~G. {et~al.} 2006, preprint (astro-ph/0603204)


\bibitem[{{Begelman} {et~al.}(1980){Begelman}, {Blandford}, \&
  {Rees}}]{Begelman:1980vb}
{Begelman}, M.~C., {Blandford}, R.~D., \& {Rees}, M.~J. 1980, \nat, 287, 307


\bibitem[{{Bekenstein}(1973)}]{Bekenstein:1973jd}
{Bekenstein}, J.~D. 1973, ApJ, 183, 657


\bibitem[{Blanchet {et~al.}(2005)Blanchet, Qusailah, \& Will}]{Blanchet:2005rj}
Blanchet, L., Qusailah, M. S.~S., \& Will, C.~M. 2005, ApJ, 635, 508


\bibitem[{Bowen \& York(1980)}]{Bowen:1980yu}
Bowen, J.~M. \& York, J.~W. 1980, Phys. Rev. D, 21, 2047


\bibitem[{{Brill} \& {Lindquist}(1963)}]{Brill:1963bl}
{Brill}, D.~R. \& {Lindquist}, R.~W. 1963, Phys. Rev., 131, 471


\bibitem[{Campanelli(2005)}]{Campanelli:2004zw}
Campanelli, M. 2005, Class. Quant. Grav., 22, S387


\bibitem[{Damour \& Gopakumar(2006)}]{Damour:2006tr}
Damour, T. \& Gopakumar, A. 2006, Phys. Rev. D, 73, 124006


\bibitem[{{Danzmann}(2003)}]{Danzmann:2003ad}
{Danzmann}, K. 2003, Advances in Space Research, 32, 1233


\bibitem[{{Dotti} {et~al.}(2006){Dotti}, {Colpi}, \& {Haardt}}]{Dotti:2006dc}
{Dotti}, M., {Colpi}, M., \& {Haardt}, F. 2006, \mnras, 367, 103


\bibitem[{Favata {et~al.}(2004)Favata, Hughes, \& Holz}]{Favata:2004wz}
Favata, M., Hughes, S.~A., \& Holz, D.~E. 2004, ApJ, 607, L5


\bibitem[{{Ferrarese}(2002)}]{Ferrarese:2002la}
{Ferrarese}, L. 2002, \apj, 578, 90


\bibitem[{{Ferrarese} \& {Merritt}(2000)}]{Ferrarese:2000fm}
{Ferrarese}, L. \& {Merritt}, D. 2000, \apjl, 539, L9


\bibitem[{{Fitchett}(1983)}]{Fitchett:1983fc}
{Fitchett}, M.~J. 1983, \mnras, 203, 1049


\bibitem[{{Fitchett} \& {Detweiler}(1984)}]{Fitchett:1984fd}
{Fitchett}, M.~J. \& {Detweiler}, S. 1984, \mnras, 211, 933


\bibitem[{Gleiser {et~al.}(1996)Gleiser, Nicasio, Price, \&
  Pullin}]{Gleiser:1996yc}
Gleiser, R.~J., Nicasio, C.~O., Price, R.~H., \& Pullin, J. 1996, Phys. Rev.
  Lett., 77, 4483

\bibitem[{Gonzalez} {et~al.}(2006){Gonzalez}, {Sperhake},
  {Bruegmann}, {Hannam}, \& {Husa}]{Gonzalez:2006md}
{Gonzalez}, J.~A., {Sperhake}, U., {Bruegmann}, B., {Hannam}, M., \& {Husa}, S.
2006, preprint (gr-qc/0610154).

\bibitem[{{Haehnelt} {et~al.}(2006){Haehnelt}, {Davies}, \&
  {Rees}}]{Haehnelt:2006hd}
{Haehnelt}, M.~G., {Davies}, M.~B., \& {Rees}, M.~J. 2006, \mnras, 366, L22


\bibitem[{{Haiman}(2004)}]{Haiman:2004ve}
{Haiman}, Z. 2004, \apj, 613, 36


\bibitem[{Herrmann {et~al.}(2006)Herrmann, Shoemaker, \&
  Laguna}]{Herrmann:2006ks}
Herrmann, F., Shoemaker, D., \& Laguna, P. 2006, preprint (gr-qc/0601026)


\bibitem[{{Madau} \& {Quataert}(2004)}]{Madau:2004mq}
{Madau}, P. \& {Quataert}, E. 2004, \apjl, 606, L17


\bibitem[{{Magorrian} {et~al.}(1998)}]{Magorrian:1997hw}
{Magorrian}, J. {et~al.} 1998, ApJ, 115, 2285


\bibitem[{{Merritt} {et~al.}(2004){Merritt}, {Milosavljevi{\'c}}, {Favata},
  {Hughes}, \& {Holz}}]{Merritt:2004xa}
{Merritt}, D., {Milosavljevi{\'c}}, M., {Favata}, M., {Hughes}, S.~A., \&
  {Holz}, D.~E. 2004, \apjl, 607, L9


\bibitem[{Micic {et~al.}(2006)Micic, Abel, \& Sigurdsson}]{Micic:2006ta}
Micic, M., Abel, T., \& Sigurdsson, S. 2006, preprint (astro-ph/0609443)


\bibitem[{{Peres}(1962)}]{Peres:1962ap}
{Peres}, A. 1962, Phys. Rev., 128, 2471


\bibitem[{{Press} \& {Schechter}(1974)}]{Press:1974ps}
{Press}, W.~H. \& {Schechter}, P. 1974, \apj, 187, 425


\bibitem[{Price \& Pullin(1994)}]{Price:1994pm}
Price, R.~H. \& Pullin, J. 1994, Phys. Rev. Lett., 72, 3297


\bibitem[{{Prince}(2003)}]{Prince:2003aa}
{Prince}, T. 2003, American Astronomical Society Meeting, 202, 3701


\bibitem[{{Richstone} {et~al.}(1998)}]{Richstone:1998ea}
{Richstone}, D. {et~al.} 1998, \nat, 395, 14


\bibitem[{Sopuerta {et~al.}(2006)Sopuerta, Yunes, \& Laguna}]{Sopuerta:2006wj}
Sopuerta, C.~F., Yunes, N., \& Laguna, P. 2006, preprint (astro-ph/0608600)


\bibitem[{{Volonteri} {et~al.}(2003){Volonteri}, {Haardt}, \&
  {Madau}}]{Volonteri:2002am}
{Volonteri}, M., {Haardt}, F., \& {Madau}, P. 2003, \apj, 582, 559


\bibitem[{Wiseman(1992)}]{Wiseman:1992dv}
Wiseman, A.~G. 1992, Phys. Rev. D, 46, 1517


\end{thebibliography}

%
\begin{deluxetable}{l c c c c c c}
\tablewidth{0pt}
\tablecaption{\label{table_fit}Non-linear fit parameters for the recoil velocity}
\tablehead{
\colhead{Model} & \colhead{$\alpha$} & \colhead{$\beta^{}_1$} & \colhead{$\beta^{}_2$}
& \colhead{$\gamma^{}_1$} & \colhead{$\gamma^{}_{2}$} & \colhead{$\gamma^{}_3$}}
\startdata
CLA                                   & $5232$  & $-2.621$ & $3.199$ & $-0.942$ & $0.808$ & $-0.405$ \\[1mm]
PN$\,(r_{1}^{\MIN})$\tablenotemark{a} & $8226$  & $0.347$  & $0.083$ & $0$      & $0$     & $0$      \\[1mm]
PN$\,(r_{2}^{\MIN})$\tablenotemark{a} & $1206$  & $0.1057$ & $0.05$  & $0$      & $0$     & $0$        
\enddata
\tablenotetext{a}{Here $r_{1}^{\MIN} = 4\,GM/c^2$ and $r_{2}^{\MIN} = 6\,GM/c^2$ are the minimum
separations up to which the PN approximation is applied.}
\end{deluxetable}
%

%
%


%
%

\end{document}